\PassOptionsToPackage{comma,authoryear}{natbib}
\documentclass[final]{l4dc2024}

\title[]{PACE: A Framework for Learning and Control in Linear Incomplete-Information Differential Games}

\usepackage{times}
\usepackage{natbib}
\usepackage{graphicx}
\usepackage{subcaption} 
\setcitestyle{authoryear,open={(},close={)},aysep={,}}
\usepackage{algorithm2e}
\usepackage{float}
\usepackage{amsmath}
\usepackage{amssymb}
\usepackage{caption}
\usepackage{hyperref}
\author{%
 \Name{Seyed Yousef Soltanian} \Email{SSOLTAN2@ASU.EDU}\\
 \addr Department of Mechanical and Aerospace Engineering, Arizona State University, Tempe, AZ, 85287, USA
 \AND
 \Name{Wenlong Zhang} \Email{WENLONG.ZHANG@ASU.EDU}\\
 \addr School of Manufacturing Systems and Networks, Arizona State University, Mesa, AZ, 85212, USA.
}

\begin{document}
\maketitle

\begin{abstract}%
In this paper, we address the problem of a two-player linear quadratic differential game with incomplete information, a scenario commonly encountered in multi-agent control, human-robot interaction (HRI), and approximation methods for solving general-sum differential games.
While solutions to such linear differential games are typically obtained through coupled Riccati equations, the complexity increases when agents have incomplete information, particularly when neither is aware of the other's cost function.
To tackle this challenge, we propose a model-based Peer-Aware Cost Estimation (PACE) framework for learning the cost parameters of the other agent. In PACE, each agent treats its peer as a learning agent rather than a stationary optimal agent, models their learning dynamics, and leverages this dynamic to infer the cost function parameters of the other agent. This approach enables agents to infer each other's objective function in real time based solely on their previous state observations and dynamically adapt their control policies. Furthermore, we provide a theoretical guarantee for the convergence of parameter estimation and the stability of system states in PACE. Additionally, in our numerical studies, we demonstrate how modeling the learning dynamics of the other agent benefits PACE, compared to approaches that approximate the other agent as having complete information, particularly in terms of stability and convergence speed.

\end{abstract}


\section{Introduction}
General-sum differential games, referring to a type of game where agents do not necessarily cooperate or compete, is a powerful tool for modeling multi-agent interactions (\citealp{schwarting2019social}) and human-robot interactions (HRI) (\citealp{li2019differential,losey2018review}).  
The feedback Nash equilibrium (FBNE) solution to general-sum non-cooperative differential games can be obtained by solving a set of coupled Hamilton-Jacobi-Isaacs (HJI) equations (\citealp{bressan2010noncooperative, crandall1983viscosity}). However, it is well known that traditional dynamic programming methods for solving these HJI equations suffer from the curse of dimensionality (\citealp{powell2007approximate}). To address this challenge, recent multi-agent research has proposed iterative linear-quadratic approximations of these games (\citealp{fridovich2020efficient}), showing the importance of studying linear-quadratic differential games.
When dealing with infinite-horizon linear quadratic differential games, it has been shown that the problem of finding the feedback Nash equilibrium can be reduced to the problem of solving a set of algebraic Riccati equations (ARE) (\citealp{bacsar1998dynamic}). Although there have been methods for solving coupled Riccati equations (\citealp{bacsar1998dynamic,cherfi2005new,li1995lyapunov}), it requires all the players to have complete information about each other's objective function parameters. The problem of finding the feedback Nash equilibrium will become challenging when dealing with agents with incomplete information. This scenario commonly arises in linear game modeling of HRI tasks when the human and the robot are not aware of each other's objective (\citealp{li2019differential,ji2018shared,franceschi2023identification, wu2024human}), or it may arise in multi-agent interactions where agents are not aware of each other's intent (\citealp{peters2021inferring,liu2016blame,li2024intent}), or in games with control imperfection (\citealp{rabbani2025optimal}). 

In this paper, we address the problem of a two-player, general-sum linear game with a quadratic infinite horizon cost under incomplete information. Our proposed Peer Aware Cost Estimation (PACE) algorithm leverages prior knowledge of the other agent's learning dynamics, enabling both agents to learn each other’s objective function in real-time while simultaneously updating their control policies based on these updated estimates. Each agent uses a history of past state observations (without any requirement to observe the other agent's actions) to minimize the error between the expected trajectory and the actual observed trajectory by updating its belief over the other agent's cost parameters using gradient descent. Our method presents a major departure from common approximation methods where an agent models the other agents as complete information or experts (\citealp{le2021lucidgames,schwarting2019social}). Despite the empirical success of these approximations in many scenarios, these methods can fail when dealing with two learning agents in certain scenarios (\citealp{liu2016blame}), as shown in Section \ref{experiments}. The core idea of PACE, however, lies in the fact that each agent simulates the learning process of the other agent at each belief update stage. As a result, avoiding biased estimates at each step rather than treating the other agent as a complete information agent. The structure of PACE allows us to theoretically guarantee the convergence of the online cost parameter inference of each agent to a bounded region around the true parameters while simultaneously ensuring the stability of the system states when both agents are performing PACE.
Although our proposed algorithm utilizes a linear time-invariant system model, its core idea can also be applied to time-varying LQ games using coupled differential Riccati equations and further extended to non-linear games through iterative LQ approximations of general multi-agent systems (\citealp{fridovich2020efficient}).

\textbf{Contributions.} \textbf{(1) Learning from an Incomplete-Information Agent.} Although differential games with multiple learners have been explored, such as in (\citealp{lian2022data}), existing work focuses on learners interacting with an expert. Our framework departs from expert-based approaches and addresses interactions between two incomplete information agents, each inferring the other's objectives. \textbf{(2) Learning from Shared State Observations.} In practical scenarios, observing another agent's actions may be impractical. Unlike works such as (\citealp{tian2023towards,li2016framework}), our algorithm relies solely on shared state observations to infer the other agent's objective. \textbf{(3) Theoretical Guarantee.} The PACE algorithm provides theoretical convergence and stability guarantees. Experimentally, we have shown its robustness and stability across diverse initial parameter estimation guesses, initial policies, and learning rates compared to the idea of treating the other agent as an expert, highlighting the importance of modeling the other agent's learning dynamics.

\section{Related Works}
\noindent\textbf{Multi-agent Interactions.} Multi-agent interactions have been extensively studied (\citealp{bloembergen2015evolutionary}) through multi-agent reinforcement learning (\citealp{canese2021multi}), adaptive control (\citealp{chen2019control}), and game theory (\citealp{yang2020overview}). Many of these interactions can be modeled as general-sum differential games (\citealp{zhang2023approximating,zhang2024pontryagin}), which are challenging to solve over continuous state and action spaces. While adaptive dynamic programming methods exist for finding feedback Nash equilibria (FBNE) (\citealp{vamvoudakis2011multi}), classical dynamic programming faces the ``curse of dimensionality'' (\citealp{powell2007approximate}). Recent work has explored iterative linear-quadratic approximations for multi-agent interactions (\citealp{fridovich2020efficient}). Under linear system and quadratic cost assumptions, Nash equilibria have been well studied (\citealp{possieri2015algebraic, engwerda1998scalar}) within the linear-quadratic non-cooperative differential game (LQR) (\citealp{bacsar1998dynamic, lukes1971global}). 
Our model-based (LQR) game approach simplifies the challenge of finding value functions in general-sum differential games by focusing on inferring the peer agent's objective function. This enables us to study the benefits of modeling the learning dynamics of other agents in multi-agent interaction tasks.

\noindent\textbf{Differential Games with Incomplete Information.} 
The existence of value and player policies for general-sum differential games with incomplete information remains an open question, unlike their zero-sum or discrete-time counterparts (\citealp{aumann1995repeated,cardaliaguet2012games}). Solving these games often requires simplifications, such as treating one agent as an uncertain learner and the others as fully informed (\citealp{le2021lucidgames,laine2021multi}), updating via belief space (\citealp{chen2021shall}), learning from offline dataset (\citealp{mehr2023maximum}), immediate cost minimization (\citealp{liu2016blame}), or using observer-based adaptive control to estimate other agents' parameters (\citealp{li2019differential,lin2023composite,franceschi2023identification}) and still majority of them focusing on one uncertain agent learning from the expert agent. Our work is motivated by the fact that incomplete information differential games, where all agents are treated as learners, have rarely been analyzed (\citealp{jacq2019learning}), even for linear systems. One example of such a study can be found in (\citealp{liu2016blame}), which compares the drawbacks of approximating the other agent as an expert versus treating them as a learner, but only for games focused on immediate cost minimization.

\noindent\textbf{Inverse Optimal Control and Inverse Reinforcement Learning.}
Traditional IOC and IRL methods rely on observing optimal trajectories to retrieve cost or reward parameters but are unsuitable for online learning and control due to their need for full trajectory observations (\citealp{molloy2022inverse,ng2000algorithms}). This has led to the development of online IOC and IRL (\citealp{molloy2020online,rhinehart2017first,self2022model}) and specifically inverse LQR (\citealp{priess2014solutions,zhang2024inverse,xue2021inverse}). Recent work on inverse non-cooperative dynamic games focuses on recovering cost functions by observing the Nash equilibria (\citealp{molloy2022inverse}), particularly for linear quadratic dynamic games (\citealp{inga2019solution,yu2022inverse,li2023cost}).
Another research addresses the problem of controlling a learner system based on observing the Nash equilibrium of an expert system and recovering their cost parameters (\citealp{lian2022data}). There has been less focus on learning from another learner (\citealp{jacq2019learning,foerster2017learning}), and not many studies, to our knowledge, explore two agents learning each other's objective function and interacting simultaneously (\citealp{liu2016blame}).
\section{Preliminaries}
\noindent\textbf{Problem Statement.} We consider a continuous-time differential game between two agents \( i \) and \( j \) interacting over an infinite horizon. The system dynamics are given by:
\begin{equation}
\label{eq1}
\dot{x}(t) = A x(t) + B_i u_i(t) + B_j u_j(t), \quad x(0) = x_0,
\end{equation}
where \( x(t) \in \mathbb{R}^n \) is the state vector, and \( u_i(t), u_j(t) \in \mathbb{R}^m \) are the control inputs of agents \( i \) and \( j \), respectively. The system matrices \( A \in \mathbb{R}^{n \times n} \), \( B_i, B_j \in \mathbb{R}^{n \times m} \) are known to both agents. Each agent aims to minimize its own infinite-horizon cost function. For agent \( k \in \{i, j\} \), the cost function is defined as:
\begin{equation}
J_k(u_k, u_{-k}) = \int_0^\infty \left( x(t)^\top Q_k x(t) + u_k(t)^\top R_k u_k(t) \right) dt,
\end{equation}
where \( Q_k \in \mathbb{R}^{n \times n} \) is a positive semi-definite state weighting matrix, \( R_k \in \mathbb{R}^{m \times m} \) is a positive definite control weighting matrix, and \( u_{-k}(t) \) denotes the control policy of the other agent.

We assume that all states \( x(t)\) are observable by both agents, but the agents are not able to observe the actions of their peer \(u_{-k}(t)\). For each agent \( k \in \{i, j\} \), the pair \( (A, B_k) \) is controllable, and the pair \( (A, \sqrt{Q_k}) \) is detectable. Notation: \( \hat{\cdot} \) denotes estimations (e.g., \( \hat{Q}_{-k} \) is agent \( k \)’s estimate of the other’s cost and \( \hat{Q}_{k} \) is agent \( k \) estimation of agent \( -k \)'s estimation of \( {Q}_{k} \)), while subscripts \( k \) and \( -k \) refer to agent \( k \) and the opposing agent, respectively. For the sake of brevity, the operator \(\mathcal{R}\mathcal{I}\mathcal{C}_k(P_k,P_{-k}, Q_k)\) represents an algebraic Riccati equation from agent \(k\)'s perspective to find \(P_k\) with a coupling term containing \(P_{-k}\) and the state cost matrix of \(Q_k\). In this work, for the sake of brevity, we assume \( R_k = R_{-k} = I \) for the rest of this paper and focus on learning the \(Q_k\) matrices.
\\
\noindent\textbf{Nash Equilibrium.} The objective of such games is to find the feedback Nash equilibrium policies \( u_i^*(t), u_j^*(t) \) such that, for each agent \( k \):
\begin{equation}
J_k(u_k^*, u_{-k}^*) \leq J_k(u_k, u_{-k}^*), \quad \forall u_k.
\end{equation}

Under the assumption of linear dynamics and quadratic costs, the value function of each agent \( k \) takes the quadratic form \(V_k^*(x) = x^\top P_k^* x\).
where \( P_k^* \in \mathbb{R}^{n \times n} \) is a positive semi-definite matrix to be determined.\\
\noindent\textbf{Coupled HJB Equations.} The Hamilton-Jacobi-Bellman (HJB) equation for agent \( k \) is:
\begin{equation}
0 = \min_{u_k} \left\{ x^\top Q_k x + u_k^\top R_k u_k + \left( \frac{\partial V_k^*}{\partial x} \right)^\top \left( A x + B_i u_i + B_j u_j \right) \right\}.
\end{equation}

\noindent\textbf{Coupled Riccati Equations.} Solving the above equation for each agent at the Nash equilibrium results in a linear control policy of \(u_{k}^* = - K_k^*x = -R_{k}^{-1}B_kP_k^*x\). Substituting this into the HJB equation and noting that \( \frac{\partial V_k^*}{\partial x} = 2 P_k^* x \), we obtain the coupled Algebraic Riccati Equations (AREs) for agents \( i \) and \( j \):
\begin{eqnarray}
0 = \left( A - B_j R_j^{-1} B_j^\top P_j^* \right)^\top P_i^* + P_i^* \left( A - B_j R_j^{-1} B_j^\top P_j^* \right) - P_i^* B_i R_i^{-1} B_i^\top P_i^* + Q_i,\\
0 = \left( A - B_i R_i^{-1} B_i^\top P_i^* \right)^\top P_j^* + P_j^* \left( A - B_i R_i^{-1} B_i^\top P_i^* \right) - P_j^* B_j R_j^{-1} B_j^\top P_j^* + Q_j,
\end{eqnarray}
which can be written as \(\mathcal{R}\mathcal{I}\mathcal{C}_k(P_{j}^*, Q_i) = 0\) and \(\mathcal{R}\mathcal{I}\mathcal{C}_k(P_{i}^*, Q_j) = 0\). These equations are coupled due to the presence of \( P_{-k}^* \) in each agent's equation. Solving these AREs yields the Nash equilibrium solutions \( P_i^*, P_j^* \), and the resultant optimal policies.

\section{PACE for Solving Incomplete Information Linear Differential Games} \label{sec:incomplete_information_game}
In scenarios where agents are unaware of each other's cost matrices \( Q_{-k} \), they need to estimate \( \hat{Q}_{-k} \) and subsequently the other agent's Riccati solution \( P_{-k} \) in their own Riccati equation as \( \hat{P}_{-k} \). We introduce PACE as a two-step algorithm: first, we explain how each agent in PACE utilizes \( \hat{P}_{-k} \) to update their control policies dynamically; next, we detail how \( \hat{P}_{-k} \) and \( \hat{Q}_{-k} \) are updated and learned during the interaction based on modeling the learning dynamic of the other agent and previous state observations.

\begin{figure}[h]
    \centering
    \includegraphics[width=1\textwidth, height=0.45\textwidth]{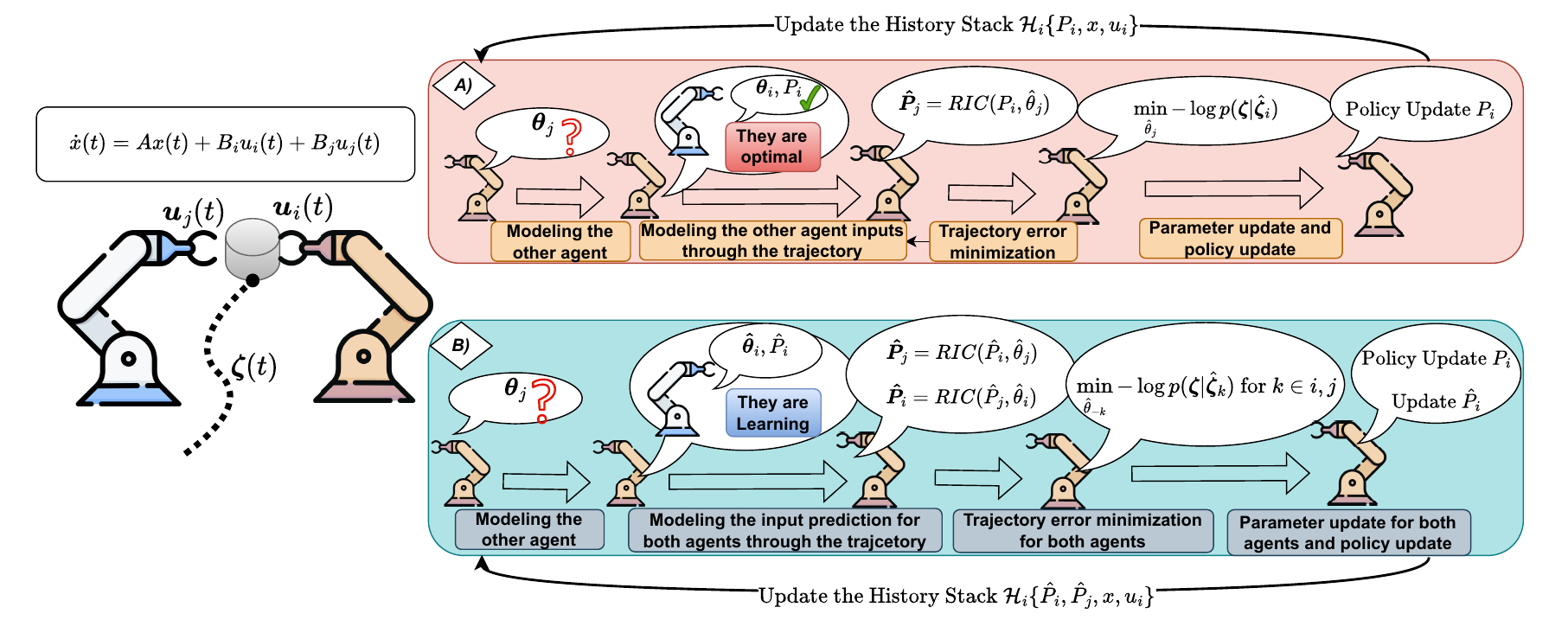}
    \captionsetup{font=small}
    \caption{An illustrative example of two robotic agents moving an object with full state trajectory observation $x(t)$, although agents are not able to observe each other's interaction force $u(t)$. Assuming an accurate low-level control of the end effectors in the task space, the interaction dynamics is modeled as a linear system. The agents are unaware of each other’s cost function parameter, denoted as $\theta$. The agent \(i\) focuses on minimizing the observed trajectory error and updating the parameter estimates in real time. However, in (A), \(i\) assumes its partner has complete information (resulting in a biased estimation), whereas in (B), \(i\) not only performs its own parameter estimation but also accounts for its partner’s learning process.}
    \label{fig:figure1}
\end{figure}

\subsection*{PACE: Policy Update} 

In PACE, at each decision-making time step \(t\) when agent \(k\) updates their estimation of the other agent cost parameters \( \hat{Q}_{-k}^{(t)}\) and \( \hat{P}_{-k}^{(t)}\), they use the following Riccati equation to update their \( {P}_{k}^{(t)}\) and their corresponding policy \(u_k(t) = -B_k^\top P_k^{(t)}x(t)\):
\begin{equation}
\label{eq:blame_me_policy}
0 = \left( A - B_{-k}B_{-k}^\top \hat{P}_{-k}^{(t)} \right)^\top P_k^{(t)} + P_k^{(t)} \left( A - B_{-k}B_{-k}^\top \hat{P}_{-k}^{(t)} \right) - P_k^{(t)}B_k B_k^\top P_k^{(t)} + Q_k.
\end{equation}

To address the incomplete information scenario, it is necessary to rely on observations of past interactions (states) to estimate \( Q_{-k} \) for both agents and recover the complete information game. A common approximation assumes that each agent treats their peer as a complete information agent, meaning their actions are considered optimal (\cite{laine2021multi}). Consequently, each agent attempts to infer the peer’s objective, as illustrated in Fig.~\ref{fig:figure1}(A). However, when dealing with two learner agents, this approach, in general, is a biased estimation (\citealp{liu2016blame}). We will experimentally demonstrate its potential failure in certain scenarios, as shown in Section \ref{experiments}.
In PACE, we develop an online learning algorithm to estimate \( \hat{Q}_{-k} \) and \( \hat{P}_{-k} \) by accounting for the learning dynamics of the peer agent, as conceptually illustrated in Fig.~\ref{fig:figure1}(B). 
\subsection*{PACE: Belief Update}
Our multi-agent learning algorithm is inspired by cognitive theories of human learning, which suggest that humans may update their forward model using the model's prediction error as loss functions (\citealp{schaefer2012beside}). Additionally, to develop our learning algorithms, we draw on the concept of history stacks from concurrent learning (CL) adaptive control (\citealp{kamalapurkar2017concurrent}). 

\noindent\textbf{Definition 1.} A \textbf{history stack} for agent \( k \) at time \( t \), denoted by \( \mathcal{H}_k^t\{x, u_k, \hat{P}_{-k}, \hat{P}_k\}\), is a collection of \( {x}(\cdot) \), \( {u_k}(\cdot) \), \( \hat{P}_{-k}^{(\cdot)} \), and \( \hat{P}_k^{(\cdot)} \) recorded at decision making sample times \( \tau_1 <\cdots < \tau_N \leq t \).

\noindent\textbf{Definition 2.} To define the \textbf{trajectory} in a continuous-time (CT) system, we represent it at time \( t \) as a sequence of sampled state values drawn from the history stack \( \mathcal{H}_k^t \), which contains \( x(\tau) \) at decision-making sample times up to \( t \). Thus, the trajectory \( \boldsymbol{\zeta}(t) \) is defined as \( \boldsymbol{\zeta}(t) = \{x(\tau) \mid \tau \in \{\tau_1, \tau_2, \dots, \tau_N\} \} \), where \( \{\tau_1, \tau_2, \dots, \tau_N\} \) are the same times recorded in \( \mathcal{H}_k^t \).

In our framework, PACE, each agent \( k \in \{i, j\} \) attempts to learn its peer \(-k\)'s state cost parameters \({Q}_{-k}\) to ultimately recover the complete information game. To achieve this, each agent updates its estimate about the other agent's objective function parameter, \( \hat{Q}_{-k} \), based on the observed trajectory \(\zeta(t)\) using a learning dynamic function \( f_k \). A natural and straightforward choice for \( f_k \) is a gradient descent function, modeling the agents as gradient learners.
The approximation methods, which treat the peer agent as a complete information agent, assume that each agent only has access to its own \( f_k \), ignoring the learning dynamics of their peer, which is assumed to have complete information. In contrast, PACE models both agents as learning agents, assuming that each agent \( k \) has access to both learning dynamic functions, \( f_k \) and \( f_{-k} \). The assumption that agents have access to each other's learning dynamics or are aware of each other's initial estimates is not limiting in many multi-agent planning scenarios, particularly when the goal is to plan for known agents or robots to work together effectively. In cases where the learning dynamic of the other agent is unknown, such as in some human-robot interaction scenarios, methods like those proposed in (\citealp{tian2023towards}) can be used. These methods employ data-driven approaches, such as using transformers in (\citealp{tian2023towards}), either in real-time or offline, to learn the peer agent's learning dynamics.

\noindent\textbf{Online Cost Parameter Learning.} We define \( \hat{\theta}_{-k} = \operatorname{vec}(\hat{Q}_{-k}) \in \mathbb{R}^{n^2} \), where \( \operatorname{vec}(\cdot) \) denotes the vectorization operator. 
In general, assuming the agents are gradient learners, we model each learning function as \(\hat{\dot\theta}_{-k}^{t} = f_k(\hat{\theta}_{-k}^{t}, \mathcal{H}_k^t\{x, u_k, \hat{P}_{-k}, \hat{P}_k\} )\). 
To infer \( \hat{\theta}_{-k} \), knowing that agents cannot observe each other's control actions, we propose that agent \( k \) uses a likelihood function over the observed trajectory \( \boldsymbol{\zeta}(t) \). As a result, agent \(k\) evaluates the accuracy of its previous estimations of \(\hat{u}_{-k}\), by assuming the state dynamics evolve from \(t = \tau_1\) to the current time according to
\begin{equation}
\label{eq:blame_me_state_dynamics}
\dot{\hat{x}}_k(\tau) = A \hat{x}_k(\tau) + B_k u_k(\tau) + B_{-k} \hat{u}_{-k}(\tau), \quad \hat{x}_k(\tau_1) = x(\tau_1),
\end{equation}
where \( \hat{u}_{-k}(t) \) is computed according to the latest estimates of \(\hat{\theta}_{-k}^t\), and \(u_k(t)\) is obviously known to agent \(k\) from \( \mathcal{H}_k\). The resulting sampled estimated trajectory, \( \hat{\boldsymbol{\zeta}}_k = \{\hat{x}_k(\tau_1), \dots, \hat{x}_k(\tau_N)\} \), is used to maximize the likelihood of observing \( \boldsymbol{\zeta} \) by minimizing the negative log-likelihood:
\begin{equation}
\label{eq:likelihood_minimization}
\min_{\hat{\theta}_{-k}} -\log p(\boldsymbol{\zeta} | \hat{\boldsymbol{\zeta}}_k), \quad \text{s.t.} \quad \hat{Q}_{-k} > 0
\end{equation}

To approximate this minimization, we replace the negative log-likelihood with a squared error loss \( \mathcal{L}_k(\hat{\theta}_{-k}) = \frac{1}{N}\sum_{\tau \in \{\tau_1, \dots, \tau_N\}} \| x(\tau) - \hat{x}_k(\tau) \|^2 \), which encourages alignment between the observed trajectory \( \boldsymbol{\zeta} \) and the estimated trajectory \( \hat{\boldsymbol{\zeta}}_k \). Each element \( \hat{\theta}_{-k(i)} \) of the vector \( \hat{\theta}_{-k} \) can be updated at each decision-making step via the gradient descent method as follows
\begin{equation}
\label{eq:theta_update_self}
\hat{\dot\theta}_{-k(i)}^{(t)} = f_k^i(\hat{\theta}_{-k(i)}^{(t)}; \mathcal{H}_k^t\{x, u_k, \hat{P}_{-k}, \hat{P}_k\} ) = - \alpha \frac{\partial \mathcal{L}_k(\hat{\theta}_{-k(i)}^{(t)}; \mathcal{H}_k^t\{x, u_k, \hat{P}_{-k}, \hat{P}_k\} )}{\partial \hat{\theta}_{-k(i)}^{(t)}},
\end{equation}
where \(\alpha\) is a constant learning rate. For agent \(k\) to model the other agent's update rule, i.e., \(f_{-k}\), agent \(k\) cannot use the same equation as \eqref{eq:blame_me_state_dynamics} because they lack knowledge about \( \mathcal{H}_{-k}\), as they are unable to observe the other agent's actions at previous time steps. To address this issue, we note that agent \(k\) is aware of the fact that \(-k\) (the other agent) is also trying to minimize the log-likelihood of the observed trajectory based on their prediction (similar to \eqref{eq:likelihood_minimization}). As a result, knowing that the observed trajectory error for \(-k\) (the other agent) is due to their incorrect estimations in \(\hat{u}_k\), agent \(k\) can use the following equation to form the error dynamics for modeling the trajectory minimization of \(-k\):
\begin{equation}
\dot{{e}}_{-k}(\tau) = A {e}_{-k}(\tau) + B_k (u_k(\tau) - \hat{u}_k(\tau)), \quad {e}_{-k}(\tau_1) = 0,
\label{eq:blame_all_error_dynamic}
\end{equation}
yielding \( \mathcal{L}_{-k}(\hat{\theta}_{k}) = \frac{1}{N}\sum_{\tau \in \{\tau_1, \dots, \tau_N\}} \|e_{-k} \|^2 \). This allows agent \( k \) to simulate \(-k\)’s learning with the gradient descent update, resulting in
\begin{equation}
\label{eq:theta_update_other}
\hat{\dot\theta}_{k(i)}^{(t)} = f_{-k}^i(\hat{\theta}_{k(i)}^{(t)}; \mathcal{H}_k^t\{x, u_k, \hat{P}_{-k}, \hat{P}_k\} ) = - \alpha \frac{\partial \mathcal{L}_k(\hat{\theta}_{k(i)}^{(t)}; \mathcal{H}_k^t\{x, u_k, \hat{P}_{-k}, \hat{P}_k\} )}{\partial \hat{\theta}_{k(i)}^{(t)}}.
\end{equation}

As a result, each agent will have access to both \(f_k\) and \(f_{-k}\). We have used \(\mathcal{H}_k^t\{x, u_k, \hat{P}_{-k}, \hat{P}_k\}\) in \eqref{eq:theta_update_other} to emphasize that the information available in \(\mathcal{H}_k^t\) is sufficient to construct \(f_{-k}\) in PACE. 
\noindent\textbf{Remark 1.} If both agents agree on their initial estimates, the proposed update rules in \eqref{eq:theta_update_other} and \eqref{eq:theta_update_self} allow both agents to track and agree on the estimation pairs \((\hat{\theta}_k, \hat{\theta}_{-k})\) and \((\hat{P}_k, \hat{P}_{-k})\) at each step. This is an important properties of PACE Resulting in a centralized learning dynamic for both agents, meaning that both agents have access to the same \((\hat{\theta}_k, \hat{\theta}_{-k})\) all the time, although they are performing their estimation independently.

Despite being non-convex, the loss functions \(\mathcal{L}_{-k}(\hat{\theta}^{(t)}_{k})\) and \(\mathcal{L}_{k}(\hat{\theta}^{(t)}_{-k})\) are differentiable with respect to \(\hat{\theta}_{k}\) and \(\hat{\theta}_{-k}\), as stated in Theorem 3.2 of (\citealp{laine2023computation}). Consequently, the choice of gradient descent for the mentioned updates \eqref{eq:theta_update_other} and \eqref{eq:theta_update_self} is inspired by its demonstrated success in non-convex optimization problems (\citealp{boyd2004convex,sutskever2013importance}).

Equations \eqref{eq:theta_update_other} and \eqref{eq:theta_update_self} require computing the estimates \(\hat{u}_{-k}(t)\) and \(\hat{u}_{k}(t)\), as well as taking their derivatives. Building on the policy update shown in \eqref{eq:blame_me_policy}, we propose that agents update the estimation of the control signals at each \(t = \tau_i\) in the history as 
\(\hat{u}_{-k}(t) = -B_{-k}^\top P_{-k}(\hat{\theta}^{(t)}_{-k}, \tau_i)x(t)\) 
and 
\(\hat{u}_{k}(t) = -B_{k}^\top P_{k}(\hat{\theta}^{(t)}_{k}, \tau_i)x(t)\). 
Here, \(P_{-k}(\hat{\theta}^{(t)}_{-k}, \tau_i)\) and \(P_{k}(\hat{\theta}^(t)_{k}, \tau_i)\) can be obtained by solving the following equations for both \(k \in \{i, j\}\) at each \(\tau_i\) in the history stack \(\mathcal{H}_k^t\):
\begin{equation}
\label{eq:blame_all_riccati}
\begin{aligned}
0 &= \left( A - B_k B_k^\top \hat{P}_k^{(\tau)} \right)^\top  
P_{-k}(\hat{\theta}^{(t)}_{-k}, \tau) 
+ P_{-k}(\hat{\theta}^{(t)}_{-k}, \tau) 
\left( A - B_k B_k^\top \hat{P}_k^{(\tau)} \right) \\
&\quad - P_{-k}(\hat{\theta}^{(t)}_{-k}, \tau) B_{-k} B_{-k}^\top P_{-k}(\hat{\theta}^{(t)}_{-k}, \tau) 
+ \hat{Q}_{-k}(\hat{\theta}^{(t)}_{-k}), \quad \text{for } k \in \{i, j\}, \, \tau \in \{\tau_1, \dots, \tau_N\}.
\end{aligned}
\end{equation}

Consequently, taking derivative from the estimated control signals requires knowing the sensitivity of \( P_{-k}(\hat{\theta}^{(t)}_{-k}, \tau)\) to each \( \hat{\theta}^{(t)}_{-k(i)} \) by finding \({\partial P_{-k}(\hat{\theta}^{(t)}_{-k}, \tau)}/{\partial \hat{\theta}^{(t)}_{-k(i)}} \) through \eqref{eq:blame_all_riccati}. This sensitivity can be obtained by solving a Lyapunov equation with a state matrix \( A_k(\tau) = A - B_k B_k^\top P_k^{(\tau)} - B_{-k} B_{-k}^\top P_{-k}(\hat{\theta}^{(t)}_{-k}, \tau) \), where \( A_k(\tau) \) is a stable Hurwitz matrix (since \( P_{-k}(\hat{\theta}^{(t)}_{-k}, \tau)\) is the stabilizing solution of the Riccati equation \eqref{eq:blame_all_riccati}), ensuring that \({\partial P_{-k}(\hat{\theta}^{(t)}_{-k}, \tau)}/{\partial \hat{\theta}^{(t)}_{-k(i)}} \) exists.

\noindent\textbf{Remark 2.} Equation \eqref{eq:blame_all_riccati} represents agent \(k\)'s estimation of agent \(-k\)'s policy update equation based on \eqref{eq:blame_me_policy}. Agent \(k\) knows that \(-k\) uses \(\hat{P}_k^{(\tau)}\) in the coupling term of their Riccati equation. Under the complete information peer approximation methods, for this infinite-horizon LQ game, agent \(k\) replaces the coupling term \(\hat{P}_k(\tau)\) in \eqref{eq:blame_all_riccati} with their true matrix \(P_k^{(\tau)}\) during parameter learning, assuming the other agent \(-k\) is optimal and has complete information about \(Q_k\) and \(P_k^{(\tau)}\).

\noindent\textbf{Remark 3.} Although PACE is a framework for differential games, like many other algorithms, its implementation must be done in discrete time. As a result, Algorithm~\ref{alg:blame_all} and the subsequent discussion on parameter and policy updates are presented in a discrete-time format for more clarity.

\noindent\textbf{Prediction and Updating the History Stack.} With the updated parameters \( \hat{\theta}^{(t+1)}_{-k} \) and \( \hat{\theta}^{(t+1)}_{k} \), agents can, at the next decision-making time \(\tau_{new}\), solve a new set of coupled Riccati equations to obtain \(\hat{P}_k^{(t+1)}\) and \(\hat{P}_{-k}^{(t+1)}\). These are then used in the policy update \eqref{eq:blame_me_policy} as predictions of the other agent's gains while also updating the history stack.

\RestyleAlgo{boxruled}
\vspace{-\baselineskip} 

\begin{algorithm}[H]
\SetAlgoLined
\small 
\DontPrintSemicolon
\SetKwInOut{Initialize}{Initialize}

\vspace{0.2em} 
\noindent\hrulefill
\vspace{-0.4em} 

\caption{PACE for Agent \( k \)}
\vspace{-0.8em}
\noindent\hrulefill
\vspace{0.2em} 

\label{alg:blame_all}

\Initialize{
    \(\hat{Q}_{-k}(0)\), \(\hat{Q}_{k}(0)\) and their corresponding Riccati solutions \( \hat{P}_k^{(0)} \) and \( \hat{P}_{-k}^{(0)} \) that stabilize the closed loop system, empty history stack \( \mathcal{H}_k \)
}
\For{each time step \(t\)}{
    \textbf{1. Observe} \( x(t) \), compute \( u_k(t) = -B_k^\top P_k^{(t)} x(t) \), apply \( u_k(t) \), and update history stack by adding \( (x(t), u_k(t), \hat{P}_k^{(t)}, \hat{P}_{-k}^{(t)}) \); maintain stack size \( N \) by removing the oldest data point if necessary
    
    \textbf{2. Trajectory Generation:}
    \Indp
    \For{each \( \tau \) in \( \mathcal{H}_k \)}{
        \textbf{a.} Solve \eqref{eq:blame_all_riccati} to find \(P_{-k}(\hat{\theta}^{(t)}_{-k}, \tau)\) for both \(k \in \{i , j\}\) using \(\hat{P}_k^{(\tau)}\) and \(\hat{P}_{-k}^{(\tau)}\) from \( \mathcal{H}_k \) and the current estimates \(\hat{Q}_{-k}\) and \(\hat{Q}_{k}\)
        
        \textbf{b.} Set \( \hat{u}_{-k}(\tau) = -B_{-k}^\top P_{-k}(\hat{\theta}^{(t)}_{-k}, \tau) x(\tau) \) and \( \hat{u}_{k}(\tau) = -B_{k}^\top P_{k}(\hat{\theta}^{(t)}_{k}, \tau) x(\tau) \)
        
        \textbf{c.} Using the system dynamics \eqref{eq:blame_me_state_dynamics}, sample the point \( \hat{x}_k(\tau) \)

        \textbf{d.} Using the error dynamics \eqref{eq:blame_all_error_dynamic}, sample the point \( \hat{e}_k(\tau) \)
    }
    \Indm
     \textbf{3. Form losses} \( \mathcal{L}_k(\hat{\theta}_{-k}^{(t)}) = \frac{1}{N} \sum_{\tau \in \{\tau_1 \dots \tau_N\}} \| x(\tau) - \hat{x}_k(\tau) \|^2 \) and \( \mathcal{L}_{-k}(\hat{\theta}_{k}^{(t)}) = \frac{1}{N}\sum_{\tau \in \{\tau_1, \dots, \tau_N\}} \|e_{-k} \|^2 \)
    
    \textbf{4. Parameter Update:}
    \Indp
    \For{each \( \hat{\theta}_{-k(i)}^{(t)} \) in \( \hat{\theta}_{-k}^{(t)} \) and \( \hat{\theta}_{k(i)}^{(t)} \) in \( \hat{\theta}_{k}^{(t)} \)}{
        \textbf{a.} Compute \( \frac{\partial \mathcal{L}_k}{\partial \hat{\theta}_{-k}^i} \) and \( \frac{\partial \mathcal{L}_{-k}}{\partial \hat{\theta}_{k}^i} \) by calculating all \(\frac{\partial \hat{x}_k(\tau)}{\partial \hat{\theta}_{-k}^{(i)}}\)  and \(\frac{\partial \hat{e}_{-k}(\tau)}{\partial \hat{\theta}_{k}^{(i)}}\)  and derivation through \eqref{eq:blame_all_riccati}
        
        \textbf{b.} Update \(\hat{\theta}^{(t+1)}_{-k(i)} = \hat{\theta}^{(t)}_{-k(i)} - \alpha \frac{\partial \mathcal{L}_k}{\partial \hat{\theta}^{(t)}_{-k(i)}}\) and \(\hat{\theta}^{(t+1)}_{k} = \hat{\theta}^{(t)}_{k(i)} - \alpha \frac{\partial \mathcal{L}_{-k}}{\partial \hat{\theta}^{(t)}_{k(i)}}\)
    }
    \Indm
    \textbf{5. Prediction:} Solve coupled Riccati equations corresponding to \(\hat{\theta}^{(t+1)}_{-k}\) and \(\hat{\theta}^{(t+1)}_{k}\) to predict \(\hat{P}_{-k}^{(t+1)}\).

    \textbf{6. Policy Update:}  using  \(\hat{P}_{-k}^{(t+1)}\) update the policy by updating \( P_k^{(t+1)} \) from \eqref{eq:blame_me_policy}
}

\vspace{-0.4em} 
\noindent\hrulefill

\end{algorithm}

\begin{theorem}
If two agents begin with initial guesses for each other’s cost parameters that yield admissible policies, then under a sufficiently small learning rate \(\alpha\) and a persistently exciting system state signal in the history stack \( \mathcal{H}_k \), \textbf{Algorithm \ref{alg:blame_all} (PACE)} converges to the true cost parameters exponentially, while maintaining system stability until reaching the Nash equilibrium.

\end{theorem}
\begin{proof} 
 [See \hyperlink{Appendix.A}{Appendix.A} for the Full Proof]
\end{proof}

The convergence speed of algorithm\ref{alg:blame_all} depends on the choice of the learning rate, and the number of samples in the history stack; it can be shown that one necessary condition for PACE to converge to the true parameters is to have a persistently exciting system state signal $x(t)$ in the interval of $t\in[\tau_1,\tau_N]$. Increasing the number of observations in the history stack can improve the convergence speed. [For more discussion on the effect of learning and history stack see \hyperlink{Appendix.B}{Appendix.B}]

\section{Numerical Experiments}\label{experiments}
In this section, we evaluate PACE through two examples. First, we perform an ablation study of a shared driving task simulation to compare PACE with complete information peer approximation methods. Second, we examine multi-parameter estimation in a repetitive task involving physical interaction between a human agent and a robotic arm's end effector. Codes at: \href{https://github.com/YousefSoltanian/PACE-L4DC2025}{Github}
\noindent\subsection*{Example 1: Monte Carlo Study in Shared Steering Driving}
\begin{figure}[h]
    \centering
    \begin{minipage}{0.55\textwidth}
        \centering
        \includegraphics[width=\textwidth, height=0.63\textwidth]{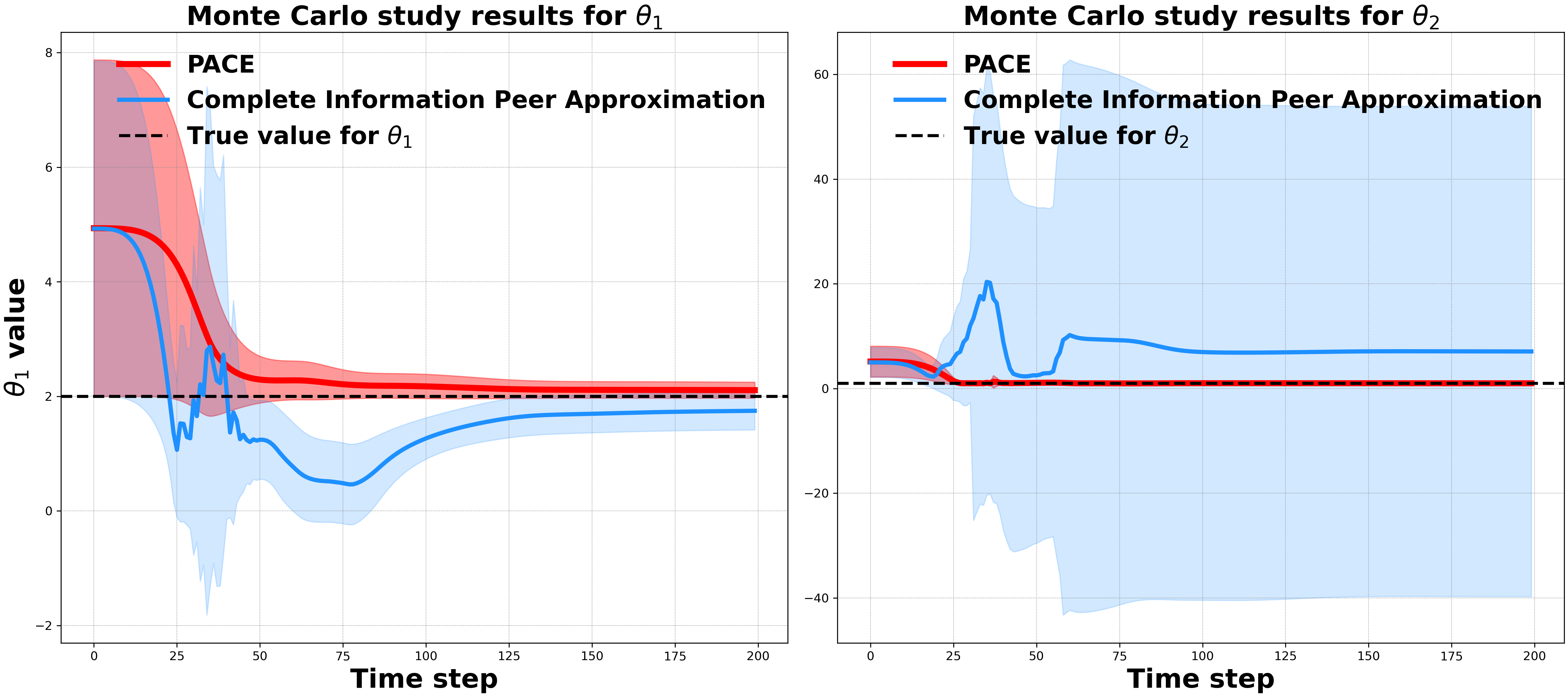}
    \end{minipage}
    \hfill
    \begin{minipage}{0.44\textwidth}
        \centering
        \includegraphics[width=\textwidth, height=0.79\textwidth]{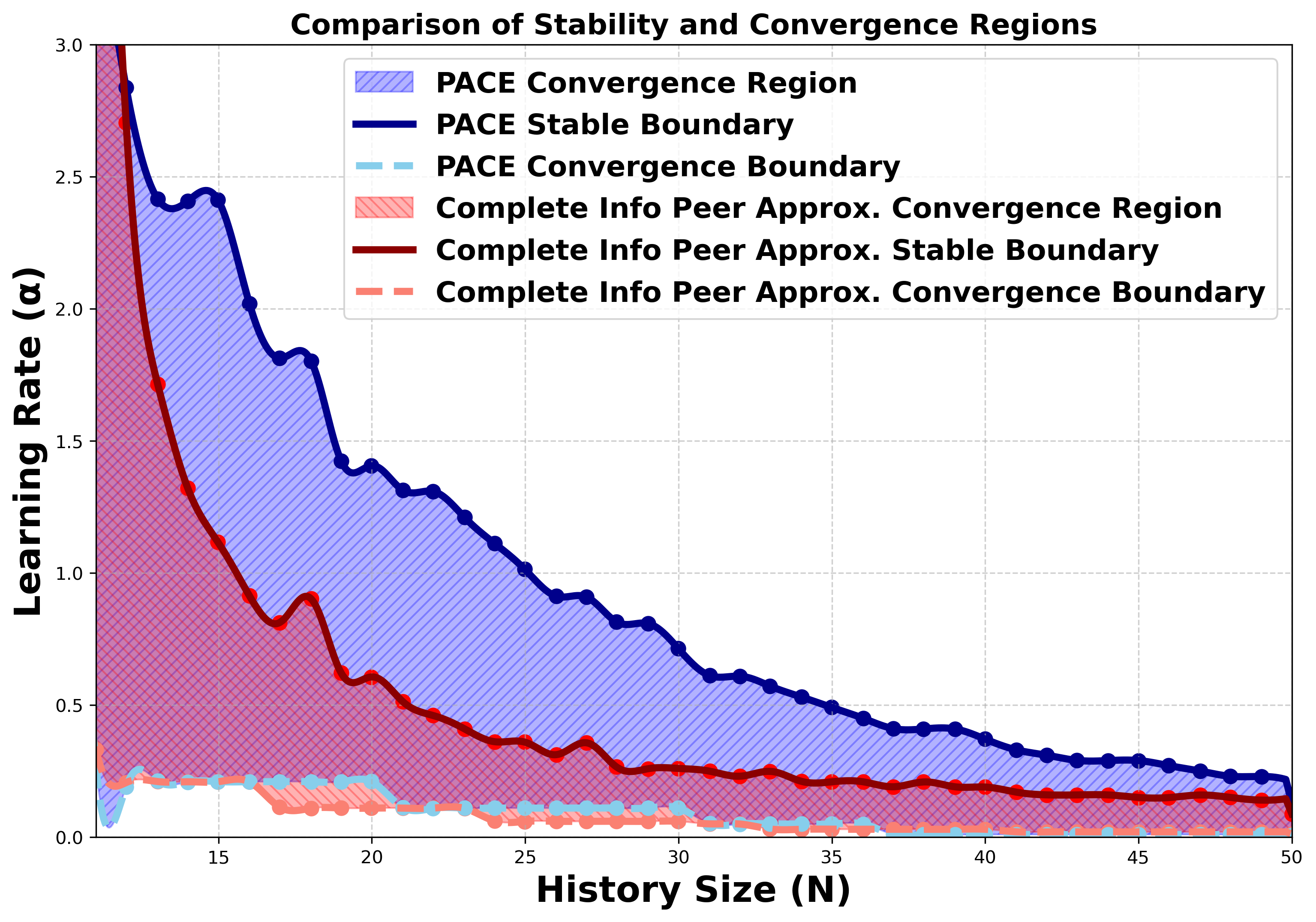}
    \end{minipage}
    \caption{Monte Carlo study results of 500 random guesses for agents' initial estimates, \( \hat{\theta}_{k}^{(0)} \) and \( \hat{\theta}_{-k}^{(0)} \) (left);  stability region analysis comparing PACE with the complete info peer approximation, showing how increasing the learning rate for each history size affect the stability boundaries(right).}
    \label{fig:exp1}
\end{figure}

The vehicle dynamics for the shared steering system is represented by the state vector \( x(t) = [e_1, \dot{e}_1, e_2, \dot{e}_2]^T \), where \( e_1 \) and \( e_2 \) denote the lateral position and orientation errors, respectively. The system dynamics is governed by the model proposed in (\citealp{rajamani2011vehicle}), with details of the matrices \( A \in \mathbb{R}^{4 \times 4} \) and \( B \in \mathbb{R}^{4 \times 1} \) available in the \hyperlink{Appendix.C}{Appendix.C} of this paper. The control inputs \( u_1(t) \) and \( u_2(t) \) represent the contributions from the human driver and the machine, respectively. The cost parameter matrices are defined as \( Q_1 = \theta_1 \cdot \text{diag}(1, 0.5, 0.5, 0.25) \) and \( Q_2 = \theta_2 \cdot \text{diag}(1, 2, 1, 0.5) \), where \(\theta_2 = 1\) and \(\theta_1 = 2\). These parameters are unknown to the other agent and need to be estimated. During the simulation, we chose a sampling time of \(0.01\) seconds.
In our first Monte Carlo study, we compared PACE with the optimal peer approximation method by modifying our learning algorithm as described in \textbf{Remark 2} (to have a fair comparison). We ran 500 simulations for both algorithms, varying the initial guesses for \(\hat{\theta}_{k}^{(0)}\) by sampling uniformly from the range \([0,10]\). The learning rate and history size are kept constant across all simulations, with \(\alpha = 0.15\) and \(N = 35\). The results are shown in Figure \ref{fig:exp1}, Left, where the system with PACE converged to at least \(80\%\) of the true values in all 500 simulations, regardless of the initial guesses. In contrast, \(85\%\) of the optimal peer approximation simulations, particularly those with larger initial errors for \(\hat{\theta}_{k}^{(0)}\), failed to converge to the true values and even became unstable during the interaction. PACE also demonstrated faster convergence, as evident in Figure \ref{fig:exp1}, highlighting its robustness and advantage over the peer approximation method.



In our second study, we fixed the initial guess \(\hat{\theta}_{k}^{(0)}\) as a near-zero random value and varied the history size \(N\) from 1 to 50. For each \(N\), we increased the learning rate incrementally, recording two critical points for each \(N\): the smallest learning rate where convergence to \(80\%\) of the true value was achieved (convergence boundary) and the learning rate at which instability occurred (stability boundary). Spline fitting visualized the convergence regions in Figure \ref{fig:exp1}, Right. The results show PACE has a wider convergence region than the optimal peer approximation method and faster convergence. Intuitively, this result makes sense as higher learning rates make agents faster learners and more nonstationary, while larger history sizes include more outdated data, undermining the complete information peer assumption and leading this method to failure.

\noindent\subsection*{Example 2: Physical Human-Robot Interaction}
\begin{figure}[h]
    \centering
\includegraphics[width=\textwidth, height = 0.315\textwidth]{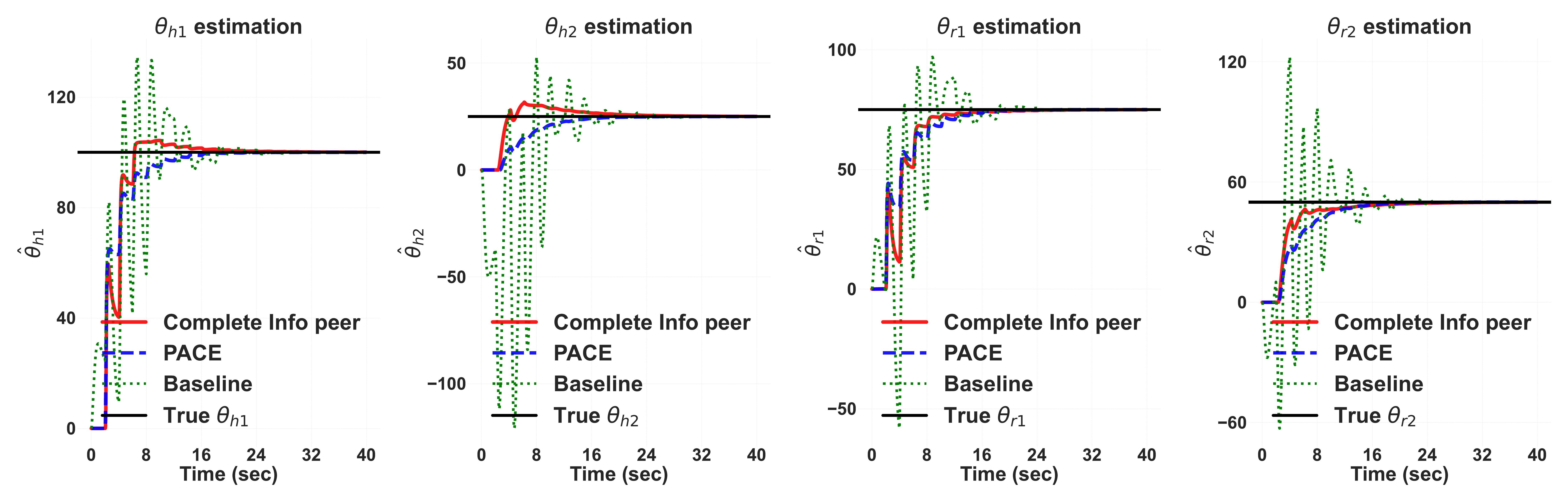}
    \caption{Comparison of three algorithms in a multi-parameter estimation scenario.}
    \label{fig:exp2}
\end{figure}

This experiment, inspired by (\citealp{li2019differential}), examines human-robot collaboration to move a robot arm's end effector between \(x_d = -10 \, \text{cm}\) and \(x_d = 10 \, \text{cm}\) every 2 seconds. The system states are \(x(t) = [e(t), \dot{e}(t)]^T\), where \(e(t) = x(t) - x_d\) is the position error and \(\dot{e}(t)\) the velocity. The system dynamics and baseline control/estimation algorithms follow (\citealp{li2019differential}), with a sampling time of \(0.01\) seconds. 
The cost matrices are \(Q_H = \text{diag}(100, 25)\) for the human and \(Q_R = \text{diag}(75, 50)\) for the robot. Unlike Example 1, this example involves multi-parameter estimation for all diagonal elements of these \(Q\) matrices. The initial state is \(x(0) = [0, 0]^T\). Both agents estimate the other's cost parameters using PACE and the complete information peer approximation, with a learning rate \(\alpha = 0.1\) and history size \(N = 15\).
Figure \ref{fig:exp2} compares PACE with the adaptive control-based method in (\citealp{li2019differential}) (baseline) and the complete information peer approximation described in \textbf{Remark 2}. PACE achieves faster convergence and reduced overshoot. Its monotonic convergence, except when \(x_d\) changes sign, is evident in Figure \ref{fig:exp2}, outperforming the baseline method. This is crucial for tasks like autonomous vehicle planning where accurate intent estimation is critical (\citealp{amatya2022shall}). [More detailed results for this experiment and its model can be found in \hyperlink{Appendix.D}{Appendix.D}]
\section{Conclusion and Future Works}
We highlighted the critical role of modeling the learning dynamics of peer agents in incomplete-information differential games through our proposed algorithm PACE. we demonstrated that PACE outperforms common approximation methods in terms of robustness, stability, monotonic convergence, theoretical guarantees, and convergence speed. While computationally intensive, PACE's applicability extends to time-varying linear systems and holds promise for adaptation to nonlinear general-sum games using iterative LQR methods in subsequent research.

 \acks{This work was supported by National Science Foundation Grants No. 1944833. The views and
conclusions contained in this document are those of the authors and should not be interpreted as
representing the official policies, either expressed or implied, of the National Science Foundation or
the U.S. Government.}


\newpage

\appendix
\section*{Appendix.A: Proof of Theorem1}
\hypertarget{Appendix.A}{}

\newtheorem{customlemma}{Lemma A.} 
\hypertarget{Appendix Al}{}
\begin{customlemma}[Monotonicity of Riccati Solutions]\label{lemma:A1}
Let \( P \) and \( \hat{P} \) be the positive definite solutions to the Riccati equations:
\begin{align}
0 &= \left( A - B_{-k}B_{-k}^\top \hat{P}_{-k} \right)^\top P + P \left( A - B_{-k}B_{-k}^\top \hat{P}_{-k} \right) - P B_k B_k^\top P + Q, \label{eq:P_riccati} \\
0 &= \left( A - B_{-k}B_{-k}^\top \hat{P}_{-k} \right)^\top \hat{P} + \hat{P} \left( A - B_{-k}B_{-k}^\top \hat{P}_{-k} \right) - \hat{P} B_k B_k^\top \hat{P} + \hat{Q}, \label{eq:hatP_riccati}
\end{align}
where \( Q \) and \( \hat{Q} \) are positive definite matrices satisfying \( Q > \hat{Q} \) (i.e., \( Q - \hat{Q} \) is positive definite). Then it follows that \( P > \hat{P} \) (i.e., \( P - \hat{P} \) is positive definite).
\end{customlemma}

\begin{proof}
Subtracting equation \eqref{eq:hatP_riccati} from \eqref{eq:P_riccati}, we obtain:
\begin{equation}\label{eq:deltaP_riccati}
0 = \left( A - B_{-k}B_{-k}^\top \hat{P}_{-k} \right)^\top \Delta P + \Delta P \left( A - B_{-k}B_{-k}^\top \hat{P}_{-k} \right) - \left( P B_k B_k^\top P - \hat{P} B_k B_k^\top \hat{P} \right) + \Delta Q,
\end{equation}
where \( \Delta P = P - \hat{P} \) and \( \Delta Q = Q - \hat{Q} \).

Since \( Q > \hat{Q} \), the term \( \Delta Q \) is positive definite. Note that \( P B_k B_k^\top P - \hat{P} B_k B_k^\top \hat{P} \) can be expressed as:
\begin{equation*}
P B_k B_k^\top P - \hat{P} B_k B_k^\top \hat{P} = \Delta P B_k B_k^\top P + \hat{P} B_k B_k^\top \Delta P.
\end{equation*}

Substituting this into \eqref{eq:deltaP_riccati}, the equation for \( \Delta P \) becomes:
\begin{equation}
0=(A-B_{-k}B_{-k}^{\top}\hat{P}_{-k}-B_{k}B_{k}^{\top}\hat{P})^{\top}\Delta P+\Delta P(A-B_{-k}B_{-k}^{\top}\hat{P}_{-k}-B_{k}B_{k}^{\top}P) + \Delta Q 
\end{equation}
This can be rearranged as:

Since \( \Delta Q > 0 \) and the closed loop system matrices \((A-B_{-k}B_{-k}^{\top}\hat{P}_{-k}-B_{k}B_{k}^{\top}\hat{P})\) and \((A-B_{-k}B_{-k}^{\top}\hat{P}_{-k}-B_{k}B_{k}^{\top}{P})\) are Hurwitz (Becuase $\hat{P}_{-k}$ and ${P}_{-k}$ are stabilizing solutions of their corresponding Riccati equations), and given the controllability assumption, the unique positive semi-definite solution of this Lyapunov-like equation is \( \Delta P > 0 \), implying \( P > \hat{P} \).
\end{proof}

\begin{customlemma}[Boundedness of the Derivative of Riccati Solution]\label{lemma:A2}
Let \( P_{-k}(\hat{\theta}_{-k}) \) be the stabilizing solution to the Riccati equation:
\[
0 = (A - B_k B_k^\top \hat{P}_k)^\top P_{-k} + P_{-k} (A - B_k B_k^\top \hat{P}_k) - P_{-k} B_{-k} B_{-k}^\top P_{-k} + \hat{Q}_{-k}(\hat{\theta}_{-k}),
\]
where \( \hat{Q}_{-k}(\hat{\theta}_{-k}) = \operatorname{vec}^{-1}(\hat{\theta}_{-k}) \). Given that \( (A, B_{-k}) \) is controllable and \( (A, \sqrt{\hat{Q}_{-k}}) \) is detectable, \( P_{-k} \) is the unique positive definite stabilizing solution. Then, for each parameter \( \hat{\theta}_{-k(i)} \) (scalar elements of the matrix \( \hat{Q}_{-k}(\hat{\theta}_{-k})\), the derivative \( \dfrac{\partial P_{-k}}{\partial \hat{\theta}_{-k(i)}} \) exists, is bounded, and satisfies:
\[
\left\| \dfrac{\partial P_{-k}}{\partial \hat{\theta}_{-k(i)}} \right\| \leq C \| S_i \|,
\]
where \( C > 0 \) is a constant dependent on the system matrices, and \( S_i \) is a matrix with a 1 at the position corresponding to \( \hat{\theta}_{-k(i)} \) and zeros elsewhere, Consequently positive semi definite.
\end{customlemma}

\begin{proof}
Differentiating the Riccati equation with respect to \( \hat{\theta}_{-k(i)} \) yields the Lyapunov equation:
\[
A_{cl}^\top \dfrac{\partial P_{-k}}{\partial \hat{\theta}_{-k(i)}} + \dfrac{\partial P_{-k}}{\partial \hat{\theta}_{-k(i)}} A_{cl} + S_i = 0,
\]
where \( A_{cl} = A - B_k B_k^\top \hat{P}_k - B_{-k} B_{-k}^\top P_{-k} \) is the closed-loop system matrix, which is Hurwitz because \( P_{-k} \) is stabilizing.

Since \( A_{cl} \) is Hurwitz, the Lyapunov equation has a unique solution, and due to the properties of the solution to a Lyapunov equation with a Hurwitz matrix (\citealp{gajic2008lyapunov}), the derivative \( \dfrac{\partial P_{-k}}{\partial \hat{\theta}_{-k(i)}} \) is bounded. Specifically, there exists \( C' > 0 \) such that:
\[
\left\| \dfrac{\partial P_{-k}}{\partial \hat{\theta}_{-k(i)}} \right\| \leq C' \| S_i \|.
\]
\end{proof}

\begin{customlemma}[Lipschitz Continuity of Riccati Solution with Respect to Parameters]\label{lemma:A3}
Let \( P_{-k}(\hat{\theta}_{-k}) \) be the stabilizing solution to the Riccati equation:
\[
0 = (A - B_k B_k^\top \hat{P}_k)^\top P_{-k} + P_{-k} (A - B_k B_k^\top \hat{P}_k) - P_{-k} B_{-k} B_{-k}^\top P_{-k} + \hat{Q}_{-k}(\hat{\theta}_{-k}),
\]
where \( \hat{Q}_{-k}(\hat{\theta}_{-k}) \) depends continuously on \( \hat{\theta}_{-k} \). Then, \( P_{-k}(\hat{\theta}_{-k}) \) is Lipschitz continuous with respect to \( \hat{\theta}_{-k} \); that is, there exists a constant \( L > 0 \) such that for any \( \hat{\theta}_{-k}, \hat{\theta}'_{-k} \),
\[
\| P_{-k}(\hat{\theta}_{-k}) - P_{-k}(\hat{\theta}'_{-k}) \| \leq L \| \hat{\theta}_{-k} - \hat{\theta}'_{-k} \|.
\]
\end{customlemma}

\begin{proof}
Since \( \hat{Q}_{-k}(\hat{\theta}_{-k}) \) is continuous in \( \hat{\theta}_{-k} \) and \( P_{-k}(\hat{\theta}_{-k}) \) is the stabilizing solution of the continuous Riccati equation, it follows that \( P_{-k}(\hat{\theta}_{-k}) \) depends continuously on \( \hat{\theta}_{-k} \).
 According to LemmaA.\ref{lemma:A2} The derivative \( \frac{\partial P_{-k}}{\partial \hat{\theta}_{-k}} \) exists and is bounded, so the mapping \( \hat{\theta}_{-k} \mapsto P_{-k}(\hat{\theta}_{-k}) \) is continuously differentiable.
Therefore, by the mean value theorem for vector-valued functions (\citealp{mcleod1965mean}), there exists a constant \( L > 0 \) such that:
\[
\| P_{-k}(\hat{\theta}_{-k}) - P_{-k}(\hat{\theta}'_{-k}) \| \leq L \| \hat{\theta}_{-k} - \hat{\theta}'_{-k} \|.
\]
This establishes the Lipschitz continuity of \( P_{-k} \) with respect to \( \hat{\theta}_{-k} \).
\end{proof}

\begin{proof}\textbf{ of Theorem.1:}
Consider the error dynamics of agent \( k \):
\begin{equation}\label{eq:error_dynamics}
\dot{e}_k(\tau) = A e_k(\tau) - B_{-k} B_{-k}^\top \tilde{P}_{-k}^{(t)}(\tau) x(\tau),
\end{equation}
where \( e_k(\tau) = x(\tau) - \hat{x}_k(\tau) \) and \( \tilde{P}_{-k}^{(t)}(\tau) = P_{-k}(\hat{\theta}_{-k}^{(t)}, \tau) - P_{-k}(\theta_{-k}, \tau) \).

Using the variation of constants formula, we can express \( e_k(\tau) \) as:
\begin{equation}\label{eq:e_integral}
e_k(\tau) = -\int_{\tau_1}^{\tau} \Phi(\tau, s) B_{-k} B_{-k}^\top \tilde{P}_{-k}^{(t)}(s) x(s) ds,
\end{equation}
where \( \Phi(\tau, s) = e^{A(\tau - s)} \) is the state transition matrix.

Then the gradient of \( e_k(\tau) \) with respect to \( \hat{\theta}_{-k}^{(t)} \) is:
\begin{equation}\label{eq:e_gradient}
\frac{\partial e_k(\tau)}{\partial \hat{\theta}_{-k}^{(t)}} = -\int_{\tau_1}^{\tau} \Phi(\tau, s) B_{-k} B_{-k}^\top \frac{\partial \tilde{P}_{-k}^{(t)}(s)}{\partial \hat{\theta}_{-k}^{(t)}} x(s) ds.
\end{equation}

Using the Lipschitz continuity of the Riccati solution with respect to \(\hat{\theta_{-k}^{(t)}}\) in \textbf{Lemma A.3} and by the Mean Value Theorem, for each \( s \in [\tau_1, \tau] \), there exists \( \omega(s) \) between \( \hat{\theta}_{-k}^{(t)} \) and \( \theta_{-k} \) such that:
\begin{equation}\label{eq:delta_P}
\tilde{P}_{-k}^{(t)}(s) = \frac{\partial P_{-k}(s)}{\partial \theta_{-k}} \bigg|_{\theta_{-k} = \omega(s)} \tilde{\theta}_{-k}^{(t)} = D_{-k}(s) \tilde{\theta}_{-k}^{(t)},
\end{equation}
where \( D_{-k}(s) \) is the Jacobian matrix of \( P_{-k}(s) \) with respect to \( \theta_{-k} \).

Similarly,
\begin{equation}\label{eq:dP_dtheta}
\frac{\partial \tilde{P}_{-k}^{(t)}(s)}{\partial \hat{\theta}_{-k}^{(t)}} = \frac{\partial P_{-k}(s)}{\partial \theta_{-k}} \bigg|_{\theta_{-k} = \hat{\theta}_{-k}^{(t)}} = D_{-k}'(s).
\end{equation}

According to the mononocity of the Riccati equation in Lemma~A.1 and from Lemmas~A.2 and~A.3, both \( D_{-k}(s) \) and \( D_{-k}'(s) \) are bounded and positive definite (PD) matrices. Substituting \eqref{eq:delta_P} into \eqref{eq:e_integral}:
\begin{equation}\label{eq:e_k}
e_k(\tau) = -\left( \int_{\tau_1}^{\tau} \Phi(\tau, s) B_{-k} B_{-k}^\top D_{-k}(s) x(s) ds \right) \tilde{\theta}_{-k}^{(t)} = -\Psi(\tau) \tilde{\theta}_{-k}^{(t)},
\end{equation}
where we define:
\begin{equation}\label{eq:Psi_def}
\Psi(\tau) = \int_{\tau_1}^{\tau} \Phi(\tau, s) B_{-k} B_{-k}^\top D_{-k}(s) x(s) ds.
\end{equation}

Similarly, substituting \eqref{eq:dP_dtheta} into \eqref{eq:e_gradient}:
\begin{equation}\label{eq:e_grad_final}
\frac{\partial e_k(\tau)}{\partial \hat{\theta}_{-k}^{(t)}} = -\int_{\tau_1}^{\tau} \Phi(\tau, s) B_{-k} B_{-k}^\top D_{-k}'(s) x(s) ds = -\Gamma(\tau),
\end{equation}
where we define:
\begin{equation}\label{eq:Gamma_def}
\Gamma(\tau) = \int_{\tau_1}^{\tau} \Phi(\tau, s) B_{-k} B_{-k}^\top D_{-k}'(s) x(s) ds.
\end{equation}
We know the loss function that is generated using past $N$ observed states in the history stack at decision-making times $\tau_n$ is in the form of:
\begin{equation}\label{eq:loss_function}
\mathcal{L}_k(\hat{\theta}_{-k}^{(t)}) = \frac{1}{N} \sum_{n=1}^{N} \| e_k(\tau_n) \|^2.
\end{equation}

Consequently the gradient of the loss function with respect to \( \hat{\theta}_{-k}^{(t)} \) is:
\begin{align}\label{eq:grad_loss}
\frac{\partial \mathcal{L}_k}{\partial \hat{\theta}_{-k}^{(t)}} &= \frac{2}{N} \sum_{n=1}^{N} e_k(\tau_n)^\top \frac{\partial e_k(\tau_n)}{\partial \hat{\theta}_{-k}^{(t)}} \nonumber \\
&= \frac{2}{N} \sum_{n=1}^{N} \left( -\Psi(\tau_n) \tilde{\theta}_{-k}^{(t)} \right)^\top \left( -\Gamma(\tau_n) \right) \nonumber \\
&= \frac{2}{N} \sum_{n=1}^{N} \tilde{\theta}_{-k}^{(t)\top} \Psi(\tau_n)^\top \Gamma(\tau_n).
\end{align}

We define:
\begin{equation}\label{eq:G_k_def}
G_k = \frac{2}{N} \sum_{n=1}^{N} \Psi(\tau_n)^\top \Gamma(\tau_n).
\end{equation}

Thus, the gradient can be written as:
\begin{equation}\label{eq:grad_loss_final}
\frac{\partial \mathcal{L}_k}{\partial \hat{\theta}_{-k}^{(t)}} = G_k \tilde{\theta}_{-k}^{(t)}.
\end{equation}

We aim to show that \( G_k \) is positive definite (PD). From Lemma~A.2 there exists $m>0$ such that for all $s,t\in[\tau_1,\tau_n]$
\[
D(s)\succeq mI,
\qquad
D'(t)\succeq mI.
\]
We define a kernel $K(s,t)$ as:
\[
K(s,t)
= D(s)\,B B^\top\,\Phi(\tau_n,s)^{T}\,\Phi(\tau_n,t)\,B B^\top\,D'(t),
\]
so we can conclude:
\[
K(s,t)\succeq m^2\,
\underbrace{B B^\top\,\Phi(\tau_n,s)^{T}\,\Phi(\tau_n,t)\,B B^\top}_{M(s,t)\succeq0}.
\]
we Define the vector:
\[
z_n \;=\;\int_{\tau_1}^{\tau_n}
\Phi(\tau_n,t)\,B B^\top\,x(t)\,dt.
\]
Then a simple rearrangement gives:
\[
\int_{\tau_1}^{\tau_n}\!\!\int_{\tau_1}^{\tau_n}
x(s)^{T}M(s,t)x(t)\,ds\,dt
=\bigl\langle z_n,z_n\bigr\rangle
=\|z_n\|^2.
\]
Hence
\[
\Psi(\tau_n)^{T}\Gamma(\tau_n)
=\int\!\!\int x(s)^T K(s,t)x(t)\,ds\,dt
\;\ge\;m^2\|z\|^2.
\]
So each $\Psi(\tau_n)^{T}\Gamma(\tau_n)$ is non-negative. By persistent excitation of $x(t)$ in the interval $t\in[\tau_1,\tau_N]$ we have $z_N\neq0$, so
$\Psi(\tau_N)^{T}\Gamma(\tau_N)>0$. Therefore
\[
v^T G_k\,v
=\frac{2}{N}\sum_{n=1}^N
\bigl(v^T\Psi(\tau_n)^T\Gamma(\tau_n)v\bigr)
>0
\quad\forall\,v\neq0,
\]
i.e.\ $G_k\succ0$.
This will conclude the convergence of the continuous time update rule to the true parameter, but because in the actual implementation, the update rule is implemented in discrete time, we continue as follow:

We consider the implementation update rule as:
\begin{equation}\label{eq:update_rule}
\hat{\theta}_{-k}^{(t+1)} = \hat{\theta}_{-k}^{(t)} - \alpha \frac{\partial \mathcal{L}_k}{\partial \hat{\theta}_{-k}^{(t)}} = \hat{\theta}_{-k}^{(t)} - \alpha G_k \tilde{\theta}_{-k}^{(t)}.
\end{equation}

Subtracting \( \theta_{-k} \) from both sides we get the estimation error dynamic:
\begin{equation}\label{eq:error_update}
\tilde{\theta}_{-k}^{(t+1)} = \tilde{\theta}_{-k}^{(t)} - \alpha G_k \tilde{\theta}_{-k}^{(t)} = (I - \alpha G_k) \tilde{\theta}_{-k}^{(t)}.
\end{equation}

Since \( G_k \) is PD, all its eigenvalues \( \lambda_i > 0 \).
Choosing the learning rate \( \alpha \) such that \( 0 < \alpha < \frac{2}{\lambda_{\max}(G_k)} \), we ensure that the spectral radius \( \rho \) of \( (I - \alpha G_k) \) satisfies \( \rho < 1 \).
Therefore, \( \tilde{\theta}_{-k}^{(t)} \) converges to zero exponentially and monotonically:
\begin{equation}\label{eq:error_convergence}
\| \tilde{\theta}_{-k}^{(t+1)} \| \leq \rho \| \tilde{\theta}_{-k}^{(t)} \|.
\end{equation}
Throughout the learning process, as the parameter estimation \(\hat{\theta}_{-k}^{(t)}\) converges to the true parameter \(\theta_{-k}\), the estimation error \(\tilde{\theta}_{-k}^{(t)} = \hat{\theta}_{-k}^{(t)} - \theta_{-k}\) approaches zero. 
Consequently, the bias in the control gain estimation \(\tilde{P}_{-k}^{(t)}(\tau) = P_{-k}(\hat{\theta}_{-k}^{(t)}, \tau) - P_{-k}(\theta_{-k}, \tau)\) diminishes for both agents. This is an important characteristic of PACE because, in PACE, by considering the learning of the peer agent, the only bias in the control gain estimation of the peer is due to the bias in the current estimated cost parameters of the learning agent.
As a result, the control gains \( P_k^{(t)}(\tau) \) and \( P_{-k}^{(t)}(\tau) \), updated using the policy update equations based on the coupling terms \( P_{-k}(\hat{\theta}_{-k}^{(t)}, \tau) \) and \( P_{k}(\hat{\theta}_{k}^{(t)}, \tau) \), get closer to the Nash equilibrium solution of the coupled Riccati equations of the complete information game.
Since the agents start with initial stabilizing admissible policies, and the control gains' estimation gets closer to their true value at each update, so they remain close to the stabilizing gains throughout the learning process, as a result, the system states remain stable during the learning.
Also, Since \( \tilde{\theta}_{-k}^{(t)} \) eventually converges to zero, as we showed, the complete information game will be recovered. As a result of the agent solving the coupled Riccati equations for estimating each other \(\hat{P}^{(t+1)}_{-k}\) and \(\hat{P}^{(t+1)}_{k}\), they will find the true Nash equilibrium solutions \(P^*_k\) and \(P^*_{-k}\) that satisfies the complete information coupled Riccati equations. So the policies will converge to Nash equilibrium policies.

\end{proof}
\section*{Appendix B: Tuning the Stepsize \(\alpha\) and History Length \(N\)}  
\label{Appendix.B}

In PACE’s discrete‐time update rule implementation in the algorithm:
\[
\tilde\theta^{(t+1)} \;=\;\bigl(I-\alpha\,G_k(N)\bigr)\,\tilde\theta^{(t)},
\]
the matrix
\[
G_k(N)\;=\;\frac{2}{N}\sum_{n=1}^N T_n,
\qquad
T_n=\Psi(\tau_n)^{T}\Gamma(\tau_n)\;\succ0,
\]
depends explicitly on the number \(N\) of samples in the history stack.  Its spectrum \(\{\lambda_i(N)\}\) governs both convergence speed and allowable \(\alpha\):

1. \textbf{Stability condition}  One requires  
\[
\rho\bigl(I-\alpha G_k(N)\bigr)<1
\;\Longleftrightarrow\;
0<\alpha<\frac{2}{\lambda_{\max}\bigl(G_k(N)\bigr)}.
\tag{B.1}\label{eq:alpha_bound}
\]
Thus the maximum admissible \(\alpha_{\max}(N)\) is
\[
\alpha_{\max}(N)
=\frac{2}{\lambda_{\max}\bigl(G_k(N)\bigr)}.
\]

2. \textbf{Dependence of \(\lambda_{\max}(G_k(N))\) on \(N\).}  
By Weyl’s inequality,
\[
\lambda_{\max}\Bigl(\sum_{n=1}^N T_n\Bigr)
\le
\sum_{n=1}^N \lambda_{\max}(T_n),
\]
so
\[
\lambda_{\max}\bigl(G_k(N)\bigr)
\le
\frac{2}{N}\sum_{n=1}^N \lambda_{\max}(T_n).
\tag{B.2}\label{eq:lammax_bound}
\]
Under the natural hypothesis that the instantaneous matrices satisfy
\(\lambda_{\max}(T_1)\le\lambda_{\max}(T_2)\le\cdots\le\lambda_{\max}(T_N)\),
the right‐hand side of \eqref{eq:lammax_bound} is a nondecreasing function of \(N\).  Consequently
\[
\lambda_{\max}\bigl(G_k(N+1)\bigr)\;\ge\;\lambda_{\max}\bigl(G_k(N)\bigr),
\]
and hence
\[
\alpha_{\max}(N+1)
=\frac{2}{\lambda_{\max}(G_k(N+1))}
\;\le\;
\frac{2}{\lambda_{\max}(G_k(N))}
=\alpha_{\max}(N).
\]
In other words, \textbf{as the history length \(N\) grows, the upper‐bound on \(\alpha\) tightens}, matching our observations in the figure (\ref{fig:exp1}) in experiment 1, that larger \(N\) requires a smaller stepsize for stability.

3. \textbf{Trade‐off with convergence rate.}  
For any admissible \(\alpha\), the asymptotic contraction factor is
\(\rho(\alpha,N)=\max_i|1-\alpha\lambda_i(N)|\), which improves (decreases) as the minimal eigenvalue \(\lambda_{\min}(G_k(N))\) grows with \(N\).  Thus adding more samples both (i) raises the intrinsic rate \(\alpha\,\lambda_{\min}(N)\) and (ii) lowers the allowable \(\alpha_{\max}(N)\).  One must choose \(\alpha\) in the interval
\[
0 \;<\;\alpha\;\le\;\alpha_{\max}(N)
\;=\;
\frac{2}{\lambda_{\max}(G_k(N))},
\]
balancing these two effects to optimize convergence.

4. \textbf{Continuous‐time limit.}  
If one instead uses the gradient‐flow
\(\dot{\tilde\theta}=-\alpha\,G_k(N)\,\tilde\theta\), then for any \(\alpha>0\) the system is exponentially stable with rate \(\alpha\,\lambda_{\min}(G_k(N))\), and no upper bound on \(\alpha\) is required.  In practice, large \(\alpha\) can amplify noise or model mismatch, but stability is guaranteed by \(G_k(N)\succ0\).

\medskip

To ensure both stability and fast convergence in discrete time, one must pick
\[
\alpha\in
\Bigl(0,\;
\frac{2}{\lambda_{\max}(G_k(N))}\Bigr],
\]
where the bound \(\lambda_{\max}(G_k(N))\) itself increases (or at least does not decrease) as \(N\) grows.  Thus empirically, larger \(N\) necessitates smaller \(\alpha\).  At the same time, increasing \(N\) raises \(\lambda_{\min}(G_k(N))\), improving the convergence rate \(\alpha\,\lambda_{\min}(G_k(N))\) for any fixed \(\alpha\) within the stability interval.  This quantifies precisely the trade‐off you observed in experiment1.

\hypertarget{Appendix.C}{}
\section*{Appendix.C: Extra Results for Example 1}
\hypertarget{Appendix.C}{}
We used the following system dynamics for our system:

 \begin{equation}
     A = \begin{bmatrix} 
     0 & 1 & 0 & 0 \\ 
     0 & -\frac{(C_f + C_r)\mu}{M v_x} & \frac{C_f + C_r}{M} & -\frac{(a C_f - b C_r)\mu}{M v_x} \\
     0 & 0 & 0 & 1 \\
     0 & -\frac{(a C_f - b C_r)\mu}{I_z v_x} & \frac{a C_f - b C_r}{I_z} & -\frac{(a^2 C_f + b^2 C_r)\mu}{I_z v_x}
     \end{bmatrix}, 
     \quad
     B_1 = 0.5B_2 = \begin{bmatrix} 0 \\ \frac{C_f \mu}{M G} \\ 0 \\ \frac{a C_f}{I_z G} \end{bmatrix}.
 \end{equation}

Here, \( M \) is the vehicle mass (1296 kg), \(v_x = 30 m/s\) is the longitudinal velocity, \( a = 1.25 \, \text{m}\) and \( b = 1.32 \, \text{m}\) are the distances from the vehicle’s center of gravity (c.g.) to the front and rear axles, respectively. \( C_f = 100700 \, \text{N/rad} \) and \( C_r = 86340 \, \text{N/rad} \) are the front and rear cornering stiffnesses, \(I_z = 1750 \, \text{kg} \cdot \text{m}^2 \) is the yaw moment of inertia, \( G = 20.46 \) is the steering ratio, and \( \mu = 0.75 \) is the road friction coefficient.

\begin{figure}[H]
    \centering
    \includegraphics[width=\textwidth]{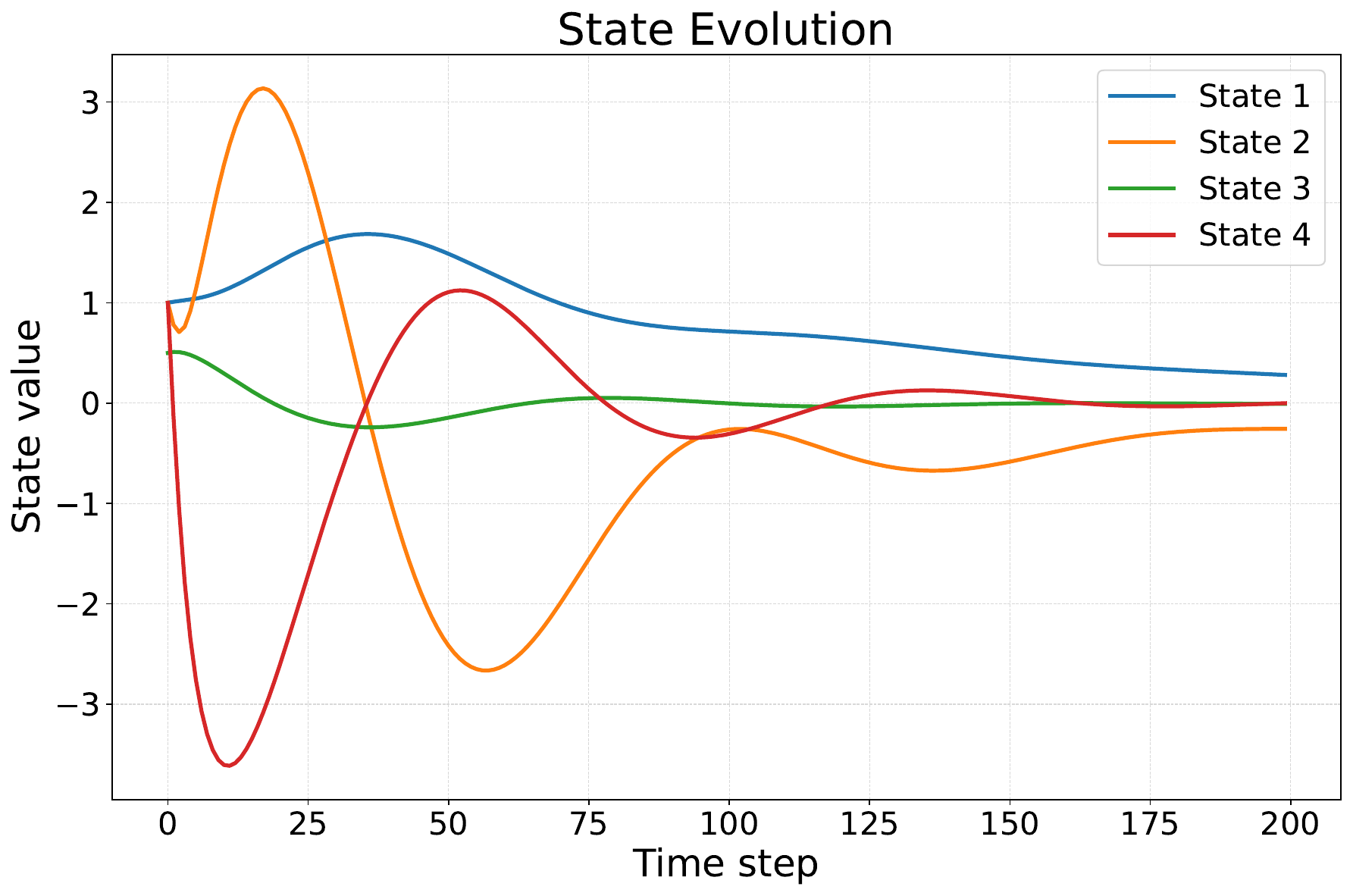}
    \caption{State evolution under the PACE algorithm. The figure shows the dynamic evolution of system states over time. Each curve represents a specific state variable's trajectory during the simulation.}
    \label{fig:AppendixB_states}
\end{figure}

\begin{figure}[H]
    \centering
    \includegraphics[width=\textwidth,  height = 0.6\textwidth]{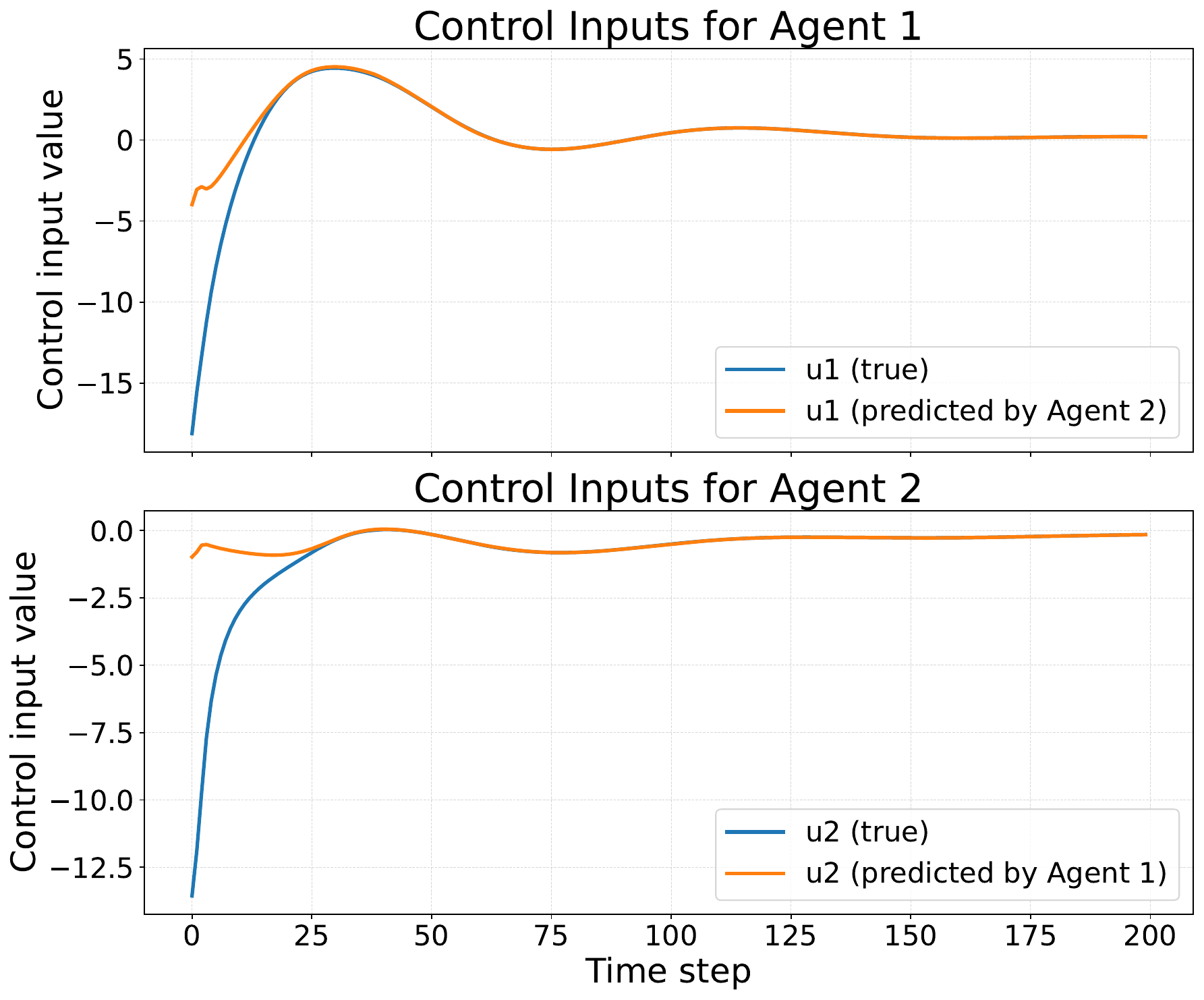}
    \caption{Control signals under the PACE algorithm. The figure compares the true control inputs (u1 and u2) with their corresponding predicted values by the other agent during the simulation.}
    \label{fig:AppendixB_control}
\end{figure}

\begin{figure}[H]
    \centering
    \includegraphics[width=\textwidth, height = 0.5\textwidth]{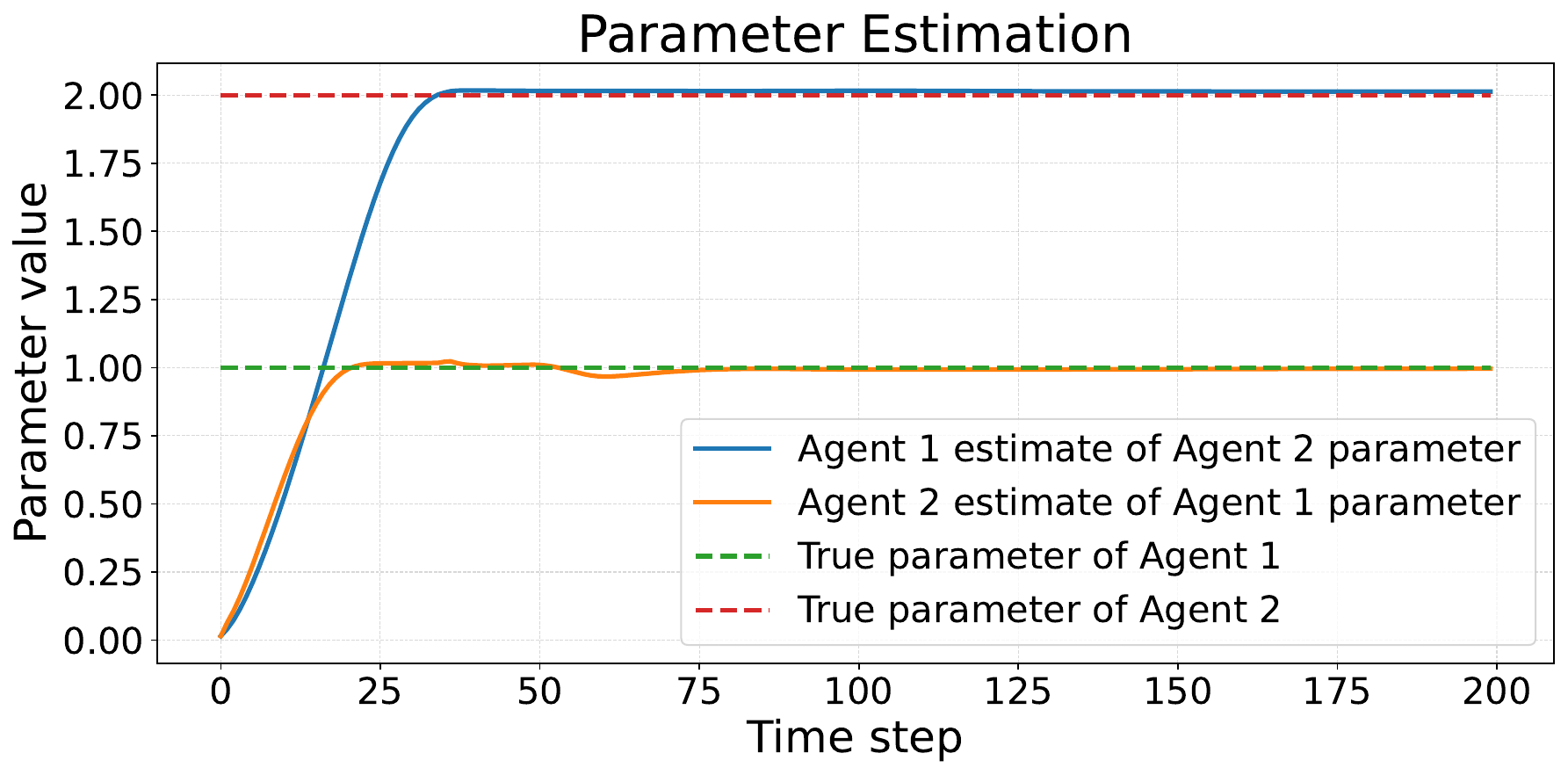}
    \caption{Parameter estimation under the PACE algorithm. The figure depicts the estimated values of cost parameters by both agents compared to their true values over time.}
    \label{fig:AppendixB_parameters}
\end{figure}

\section*{Appendix.D: Extra Results for Example 2}
\hypertarget{Appendix.D}{}
The proposed system dynamic for this example is as follows:
\[
\dot{x}(t) = 
\begin{bmatrix}
0 & 1 \\
0 & -\frac{C}{M}
\end{bmatrix} x(t) + 
\begin{bmatrix}
0 \\
\frac{1}{M}
\end{bmatrix} \big(u_H(t) + u_R(t)\big),
\]
where \(M = 6 \, \text{kg}\) is the mass of the robotic arm, \(C = 0.2 \, \text{Ns/m}\) is the damping coefficient, and \(x(t) = \begin{bmatrix} e(t) \\ \dot{e}(t) \end{bmatrix}\), with \(e(t)\) as the position error and \(\dot{e}(t)\) as the velocity.

Table (\ref{table2}) further compares all three algorithms on time to achieve less than 10\% error, final error values, and the standard deviation of error over the 40-second simulation, highlighting PACE's superior performance and stability.
\begin{table}[h!]
\centering
\caption{Comparison of complete info peer, PACE, and baseline estimation metrics for $\theta$ parameters}
\resizebox{\textwidth}{!}{  
\begin{tabular}{|l|ccc|ccc|ccc|}
\hline
\textbf{} & \multicolumn{3}{c|}{\textbf{Complete Info Peer approximation}} & \multicolumn{3}{c|}{\textbf{PACE}} & \multicolumn{3}{c|}{\textbf{Baseline}} \\
\textbf{Metric} & \%err std & time err $<$ 10\% & final err\% & std & time err $<$ 10\% & final error\% & std & time err $<$ 10\% & final err\% \\
\hline
\textbf{$\theta_{h1}$} & 23.67 & 6.19s & $3.93 \times 10^{-4}$ & 22.74 & 4.28s & $7.00 \times 10^{-6}$ & 22.35 & 12.14s & $8.09 \times 10^{-3}$ \\
\textbf{$\theta_{h2}$} & 25.73 & 9.29s & $2.16 \times 10^{-3}$ & 26.89 & 7.11s & $4.50 \times 10^{-5}$ & 126.34 & 19.09s & $7.55 \times 10^{-2}$ \\
\textbf{$\theta_{r1}$} & 25.69 & 8.14s & $9.61 \times 10^{-4}$ & 23.95 & 6.20s & $2.00 \times 10^{-5}$ & 32.28 & 14.02s & $8.96 \times 10^{-3}$ \\
\textbf{$\theta_{r2}$} & 24.52 & 7.08s & $8.25 \times 10^{-4}$ & 25.86 & 6.73s & $1.50 \times 10^{-5}$ & 45.93 & 15.18s & $3.29 \times 10^{-2}$ \\
\hline
\end{tabular}
}
\label{table2}
\end{table}

\begin{figure}[H]
    \centering
    \includegraphics[width=\textwidth]{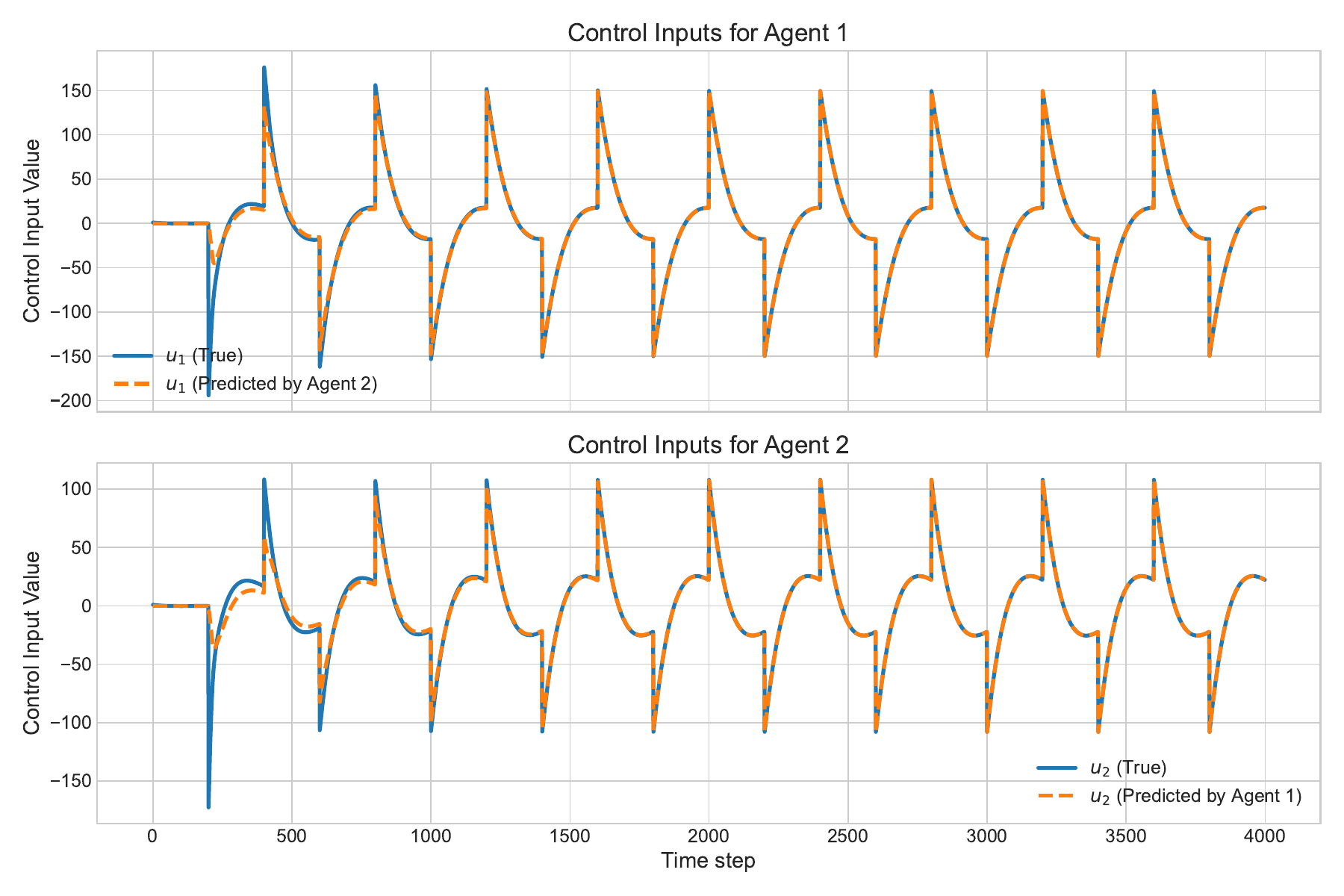}
    \caption{Control signals under the PACE algorithm. The figure compares the true control inputs (u1 and u2) with their corresponding predicted values by the other agent during the simulation for experiment 2.}
    \label{fig:AppendixC_control}
\end{figure}

\end{document}